\documentclass[twocolumn,pra,longbibliography,aps,superscriptaddress,flushbottom,showpacs]{revtex4-1}
\usepackage{xspace,amsmath,amsfonts,amsthm,amssymb,amsbsy,graphicx,color,enumerate}

\begin{document}

\newcommand{\ms}[1]{\mbox{\scriptsize #1}}
\newcommand{\msb}[1]{\mbox{\scriptsize $\mathbf{#1}$}}
\newcommand{\msi}[1]{\mbox{\scriptsize\textit{#1}}}
\newcommand{\nn}{\nonumber} 
\newcommand{\dg}{^\dagger}
\newcommand{\smallfrac}[2]{\mbox{$\frac{#1}{#2}$}}
\newcommand{\ket}[1]{| {#1} \ra}
\newcommand{\bra}[1]{\la {#1} |}
\newcommand{\pfpx}[2]{\frac{\partial #1}{\partial #2}}
\newcommand{\dfdx}[2]{\frac{d #1}{d #2}}
\newcommand{\half}{\smallfrac{1}{2}}
\newcommand{\s}{{\mathcal S}}
\newcommand{\jord}{\color{red}}
\newcommand{\kurt}{\color{blue}}
\newtheorem{theo}{Theorem} \newtheorem{lemma}{Lemma}

\title{Fermi's golden rule, the origin and breakdown of Markovian master equations, \\ and the relationship between oscillator baths and the random matrix model}

\author{Siddhartha Santra} 
\affiliation{U.S. Army Research Laboratory, Computational and Information Sciences Directorate, Adelphi, Maryland 20783, USA}
\author{Benjamin Cruikshank}
\affiliation{U.S. Army Research Laboratory, Computational and Information Sciences Directorate, Adelphi, Maryland 20783, USA} 
\affiliation{Department of Physics, University of Massachusetts at Boston, Boston, MA 02125, USA} 
\author{Radhakrishnan Balu}
\affiliation{U.S. Army Research Laboratory, Computational and Information Sciences Directorate, Adelphi, Maryland 20783, USA} 
\author{Kurt Jacobs}
\affiliation{U.S. Army Research Laboratory, Computational and Information Sciences Directorate, Adelphi, Maryland 20783, USA} 
\affiliation{Department of Physics, University of Massachusetts at Boston, Boston, MA 02125, USA} 
\affiliation{Hearne Institute for Theoretical Physics, Louisiana State University, Baton Rouge, LA 70803, USA} 

\begin{abstract}  
Fermi's golden rule applies to a situation in which a single quantum state $|\psi\rangle$ is coupled to a near-continuum. This ``quasi-continuum coupling'' structure results in a rate equation for the population of $|\psi\rangle$.  Here we show that the coupling of a quantum system to the standard model of a thermal environment, a bath of harmonic oscillators, can be decomposed into a ``cascade'' made up of the quasi-continuum coupling structures of Fermi's golden rule. This clarifies the connection between the physics of the golden rule and that of a thermal bath, and provides a non-rigorous but physically intuitive derivation of the Markovian master equation directly from the former. The exact solution to the Hamiltonian of the golden rule, known as the Bixon-Jortner model, generalized for an asymmetric spectrum, provides a window on how the evolution induced by the bath deviates from the master equation as one moves outside the Markovian regime. Our analysis also reveals the relationship between the oscillator bath and the  ``random matrix model'' (RMT) of a thermal bath. We show that the cascade structure is the one essential difference between the two models, and the lack of it prevents the RMT from generating transition rates that are independent of the initial state of the system. We suggest that the cascade structure is one of the generic elements of thermalizing many-body systems. 
\end{abstract} 

\maketitle 

\section{Introduction} 
\label{intro}

The ability to model the effects of the environment on simple quantum systems is important in a wide range of fields, including those relevant to quantum technology and quantum information. Many systems that are important in quantum technologies are subject to relatively weak coupling to environments that, as a result, behave as generic thermal baths. This coupling induces relaxation and decoherence, and can be modeled by an interaction with a near-continuum of harmonic oscillators. This model of a bath leads to a very simple ``Markovian'' master equation for the system alone so long as certain conditions are satisfied by various timescales of the system and bath~\cite{Gardiner10, Breuer07}. This master equation is essentially a set of rate equations for the populations of, and coherences between, the various discrete energy levels of the system, and is extremely useful in modeling environmental effects in a wide range of systems. 

The standard derivation of the master equation using a bath of harmonic oscillators is perturbative, involves approximations that are somewhat obscure (e.g. the approximation that the system and bath are in a product state at each time-step is demonstrably false), and provides little intuition about what happens as one moves outside the rate-equation regime~\cite{Gardiner10, Breuer07}. Nevertheless the result of this derivation is a set of valid conditions that determine in what regime such a bath will induce simple rate-equation dynamics. Here we introduce a way to analyze the dynamics of an oscillator bath that is quite different from traditional methods. Our motivations for presenting this approach, in which the interaction with the system is decomposed into a ``cascade'' of simple coupling structures, are as follows. First, it clarifies the relationship between the coupling structure of Fermi's golden rule and that of an oscillator bath. It shows that the essential mechanism for generating rate equations is the same for both and clarifies the role of various parameters in this mechanism, such as the system transition frequency and the bath temperature.  We suggest that the coupling structure of Fermi's golden rule may be the only one by which quantum mechanics is able to generate rate equations. Second, because the Hamiltonian of the golden rule can be solved (almost) exactly~\cite{Bixon68, Barnett03, Fain88a, Fain88b}, this approach provides insight into the distinct ways in which the evolution induced by an oscillator bath deviates from that of the master equation as the various conditions are relaxed. Third, it makes simple the relationship between the oscillator bath and another model of a thermal bath, the so-called ``random matrix model''~\cite{Massimiliano03, Massimiliano03b, Lebowitz04, Gemmer06, Breuer06, Bartsch08, Silvestri14}. We show that while the oscillator bath generates a master equation in which the transition rates are independent of the initial state of the system, the random matrix model cannot do so. This difference is a result of the one essential structural difference between the two models, which is that the RMT lacks the cascade structure. Our analysis suggests that this structure is a consequence of the many-body nature of thermal baths, and thus is likely to be universal. This in turn suggests that state-independent transition rates are a general feature of real thermalizing baths. 

The emergence of rate equations from quantum mechanics is also of fundamental interest from the point of view of the quantum-to-classical transition, the study of how classical dynamics arises from quantum dynamics as an emergent phenomenon~\cite{Spiller94, Percival98, Bhattacharya00, Everitt05}. Two classes of noise processes are ubiquitous in the classical world, the first is Gaussian noise and the second consists of ``jump processes'' of which the Poisson process is an example~\cite{Jacobs10c}. Gaussian noise emerges from the quantum dynamics of thermalizing (many-body) systems (a fact connected to a remarkable theorem by Kac~\cite{Kac59, Jacobs14}), while jumps processes are described by rate equations. The question of the (quantum mechanical) coupling structure(s) that are able to generate rate equations from unitary evolution is thus of fundamental interest. 

In the next section we begin by defining the (thermal) Markovian master equation for weak damping, and enumerate the conditions under which it accurately describes the effect of a bath of oscillators. In Section~\ref{struct} we show that a coupling to an oscillator bath can be decomposed into a cascade of the coupling structures of Fermi's-golden-rule (FGR).   In Section~\ref{ObME} we derive the master equation from this structure and discuss some additional implications. In Section~\ref{secFGR} we examine the exact solution for the FGR coupling structure (the Bixon-Jortner model)~\cite{Bixon68, Barnett03, Fain88a, Fain88b}, and in Section~\ref{3dev} we use this to obtain information about the distinct ways in which the dynamics of an open system deviates from the master equation as the various conditions are relaxed. In Section~\ref{RMM} we describe the ``random matrix model'' for open systems and explain how it correctly includes key elements of a thermal bath. We then show that this model cannot produce the required initial-state-independence of the master equation, and that the ability to do so is provided by the cascade structure of the oscillator bath. In Section~\ref{conc} we finish with some concluding remarks. 

\section{The thermal master equation for \\ weakly damped systems} 
\label{MEW} 

An interaction with an oscillator bath will induce a simple rate equation for a quantum system only when a number of conditions are satisfied. First there are conditions on the time-dependence of the system Hamiltonian. Here we assume that the system Hamiltonian is time independent, or if it is time-dependent then this time-dependence causes only insignificant changes to the system's eigenvalues and eigenvectors. This condition means that the dynamics induced by any time-dependent control is slow compared to the frequencies of the system's transitions (to be defined below). 
Master equations can be derived for certain classes of time-dependent systems in which the above condition is broken, but we will not deal with these here~\cite{Alicki10, Rivas10, Szczygielski13, Szczygielski14, Szczygielski15}.  

To obtain the ``standard'' Markovian master equation, there are two remaining conditions that involve the parameters of the system. (Further conditions that involve the parameters of the bath itself will be discussed in the next section.) These conditions are required by the rotating-wave approximation (RWA)~\cite{Jacobs14, Breuer07, Davies1974, Redfield57, Redfield65}, and are also referred to as the conditions for \textit{weak damping}. First we need to   introduce some terminology. Let us denote the energy eigenstates of our quantum system by $|\psi_n\rangle$, $n=0,\dots, N-1$, their respective energies by $E_n$ and their respective frequencies by $\tilde{\Omega}_n = E_n/\hbar$. A transition is defined as a pair of states, $\{ |\psi_n\rangle, |\psi_m\rangle \}$, ordered so that $E_n > E_m$. We will label all the transitions of a system by $\mbox{T}_j$ with $j = 1,\ldots, M$. The ``frequency'' of a system transition $j$ is $\Omega_j \equiv \tilde{\Omega}_n - \tilde{\Omega}_m$. The ``damping rate'' of a system transition, $\gamma_j$, is the total rate at which jumps from $|\psi_n\rangle$ to $|\psi_m\rangle$ are induced by the bath when the bath is at zero temperature. The actual rate at which jumps occur for any given transition depends not only on $\gamma_j$ but also on the temperature $T$. In particular, the rate at which downward jumps occur is $\gamma_j^{-} = \gamma_j (n_T[\Omega_j] + 1)$. Here $n_T[\Omega_j]$, given below in Eq.(\ref{nT}), is the average number of quanta in a harmonic oscillator that has frequency $\Omega_j$ and is at temperature $T$. The rate at which upward jumps occur is $\gamma_j^{+} =  \gamma_j  n_T[\Omega_j]$. 

A system is weakly damped if 
\begin{equation}
\begin{array}{l}
\;\, \mbox{1.}  \;\;\;\;\;\;\;\;   \gamma_j \ll \Omega_j , \;\;\;  \forall j ,    \\
    \\ 
    \left. \begin{array}{rc} \mbox{2.}  \;\;\; \mbox{Either i)}  & \; |\Omega_k - \Omega_j| \gg \max (\gamma_j,\gamma_k)   \\
                                                   \mbox{or  ii)} & \; |\Omega_k - \Omega_j| \ll \min (\gamma_j,\gamma_k) \end{array}  \right\}  \;\;\forall j,k 
\end{array}
\end{equation}
Condition 2 ensures that all pairs of transitions can be separated into those that are degenerate (have the same transition frequency from the point of view of the system/bath coupling) and those that are non-degenerate with respect to this coupling. If there are two or more transitions that are degenerate it is useful to re-label the transitions by the distinct frequencies. We will denote each distinct transition frequency by $\nu_j$, the number of transitions that are effectively degenerate with this frequency by $K_j$, and label each of the transitions by $\mbox{T}_{jk}$, $k = 1, \ldots, K_j-1$. Of course each of the transitions has an upper state and a lower state. We will denote the upper state by $|\mbox{e}_{jk}\rangle$ and the lower state by $|\mbox{g}_{jk}\rangle$. We present this notation in Table~\ref{tabsyms1} for convenience. 

\begin{table}[t]
\caption{Notation to incorporate degenerate transitions}
\begin{center}
\begin{tabular}{ccl}
  Symbol  & & Meaning \\ 
  \hline
   $\nu_j$ & & Distinct transition frequencies \\
 $K_j$ & & Degeneracy of  $\nu_j$ \\
 T$_{jk}$ & & $k^{\msi{th}}$ transition with frequency $\nu_j$  \\
 $L_j$ & & Transition operator for frequency  $\omega_j$ \\
 $|\mbox{e}_{jk}\rangle$ & & Upper state for transition T$_{jk}$ \\
 $|\mbox{g}_{jk}\rangle$ & & Lower state for transition T$_{jk}$ 
\end{tabular}
\end{center}
\label{tabsyms1}
\vspace{-0.4cm}
\end{table}%

We can write the thermal Markovian master equation in a compact form by using the definitions given in Table~\ref{tabsyms1} and define ``transition operators'', $L_j$, by   
\begin{align}
  L_{j} = \sum_{k=1}^{K_j} |\mbox{g}_{jk}\rangle \langle \mbox{e}_{jk}| . 
\end{align}
Note that each $L_j$ describes jumps for all the transitions that share the transition frequency $\nu_j$. The transitions arise from the coupling to the bath, a coupling that is proportional to some system operator $X$. There will be transitions between states $|\mbox{g}_{jk}\rangle$ and $|\mbox{e}_{jk}\rangle$ only if $X$ has a non-zero matrix element between these states. We also define a superoperator $\mathcal{D}$ by  
\begin{align}
   \mathcal{D}[c] \rho & \equiv c \rho c^\dagger -  \left( c^\dagger c \rho + \rho c^\dagger c \right)/ 2 
\end{align}
for any operator $c$, and the quantity 
\begin{align} 
  n_T[\omega]  =  (e^{\hbar\omega/kT} - 1)^{-1}   
  \label{nT}
\end{align}
in which $k$ is Boltzmann's constant. This quantity is the average number of quanta in a harmonic oscillator with frequency $\omega$ at temperature $T$. We can now write the thermal Markovian master equation for weakly damped systems, which is 
\begin{align} 
   \dot{\rho} = & -\frac{i}{\hbar} [H + H_{\ms{L}},\rho] +  \biggl[ \sum_j \gamma_j \mathcal{D}[L_j] \biggr] \rho  \nonumber \\ 
   & +  \sum_j \gamma_j \, n_T[\omega_j] \, \biggl[ \mathcal{D}[L_j]  + \mathcal{D}[L_j^\dagger]  \biggr]  \rho . 
   \label{MME}
\end{align} 
This equation tends to be referred to as the ``standard'' master equation; it describes the thermal relaxation of quantum systems in terms of a simple (Markovian) rate dynamics~\cite{Davies1974, Breuer07}. It reduces, for example, to the ``quantum optical'' master equation for a cavity mode damped by a lossy end-mirror~\cite{Jacobs14}. This master equation, while relevant in many applications, appears to lack a concise and convenient name. For want of such a name, we will refer to it here as the ``MEW'', an acronym drawn from the phrase ``master equation for weak damping''. 

In the above equation $H_{\ms{L}}$ gives the (small) changes to the energy levels due to the Lamb shift. These shifts depend explicitly on how the coupling between the system and the bath oscillators tapers off at high frequencies. We may choose this coupling to end abruptly at some sufficiently high frequency $\Omega$. In this case the bath is referred to as having a ``sharp cut-off'' and $\Omega$ is called the \textit{cut-off frequency} of the bath. We can alternatively allow the coupling to the bath oscillators to reduce gradually to zero in some way above some designated frequency, which in that case is also referred to as the ``cut-off''. The Lamb shifts of the energy levels depend not only on the cut-off frequency but also on the nature of the cut-off. Here we will always use a sharp cut-off at a frequency $\Omega_{\ms{c}}$. We discuss the origin and calculation of the Lamb shift in Section~\ref{shift}.  

The standard derivation of the MEW, Eq.(\ref{MME}), can be found in~\cite{Breuer07, Alicki89, Davies1974, Redfield57, Redfield65}. Note that the MEW is derived from the Schr\"{o}dinger equation for the system and bath combined, and this equation is deterministic. Quantum mechanics however has randomness built into the definition of the states. The MEW, while itself deterministic, can be thought of as describing a random process, in a way very similar to the classical master equations that describe dynamics in which random jumps are occurring between discrete states~\cite{JacobsSP}. Such classical master equations are deterministic only because they describe the evolving probability distribution over the states of the system, as opposed to an actual sample evolution in which the system will be seen to jump between states at random times. Similarly the dynamics described by the MEW can be rewritten as a stochastic equation for system in a pure state that changes at random times. These stochastic evolutions for quantum systems have been termed ``quantum trajectories''~\cite{Carmichael89}, or ``quantum Monte Carlo''/``quantum jump'' methods''~\cite{Molmer93, Hegerfeldt92, Srinivas81}, but we will not be concerned with them here.  

\section{The structure of an oscillator bath} 
\label{struct}

An oscillator bath consists of a large number of oscillators whose frequencies are closely spaced on the real line. So long as the spacing of the oscillators is much smaller than all frequencies of the system, including the inverse of the time over which we wish to evolve the system, then the bath is well characterized by a quantity called the \textit{spectral density}, usually denoted by $J(\omega)$. This density is the product of the actual density per unit frequency of the oscillators (assuming the limit in which the spacing between the oscillators tends to zero) with the square of the coupling between the system and each oscillator, which may also vary with frequency. Specifically, if we denote the annihilation operator for the oscillator with frequency $\omega$ by $a_\omega$, and write the Hamiltonian of the system and oscillator bath by 
\begin{equation} 
    H = H_0 +  \hbar X \!\! \int_0^\infty \!\!\!\! \tilde{g}(\omega)\, x_\omega \, d\omega + \! \int_0^\infty \!\!\!\! \tilde{D}(\omega)\, a_\omega^\dagger a_\omega \, d\omega , 
\end{equation}
then the spectral density is 
\begin{equation} 
    J(\omega) = \tilde{g}^2(\omega) \tilde{D}(\omega) . 
\end{equation}
Here $H_0$ is the Hamiltonian of the system, $X$ is a dimensionless system observable, and $x_\omega = a_\omega + a_\omega^\dagger$. Note that we have defined the coupling ``strength'' $\tilde{g}$, oscillator density $\tilde{D}$, and $J$ so that they are all dimensionless. We can define a version of the oscillator density, $D(\omega)$, that actually has units of inverse frequency by pulling out an (arbitrary) frequency, $\omega_d$, from the integral so as to write $\tilde{D}(\omega) d\omega = \omega_d D(\omega) d\omega$. 

For weak coupling, as we shall see in what follows, each system transition couples primarily only to the oscillators within a small frequency band around the transition frequency. The resulting transition rates are then determined, to good approximation, by the value of the spectral density at the frequency of each transition. Thus it is only the values of the spectral density at each of the transition frequencies that affects the master equation, and this effect is merely to scale up or down each of the transition rates. Because the spectral density affects the resulting master equation only in this simple way, it is not important for our analysis exactly how we chose this density, except that we must consider what happens to it as $\omega \rightarrow \infty$, as we now explain. 

The coupling strength $\tilde{g}(\omega)$ tends to zero as $\omega$ tends to infinity due to the fact that the spatial extent of the system is nonzero. For example, the effect on a given charge distribution of variations in the electric field that are much smaller than the volume of the distribution average themselves out over the volume. A system therefore has a ``cut-off'' frequency above which it does not couple to the bath oscillators. For our analysis we will take the bath to have a ``sharp'' cut-off frequency, $\Omega_{\ms{c}}$, meaning that $\tilde{g}(\omega)$ is non-zero up until $\Omega_{\ms{c}}$, and zero for all higher frequencies. We will also choose the simplest form for the coupling strength $\tilde{g}$ and the oscillator density, $D(\omega)$, setting them to be constant between $\omega = 0$ and $\omega = \Omega_{\ms{c}}$. In particular $D(\omega) = D$ for $\omega \in [0, \Omega_{\ms{c}}]$. 

Note that the bath actually consists of a \textit{discrete} set of oscillators --- the continuum description is merely an idealization. Choosing the spectral density of the oscillators to be uniform naturally means that the oscillators are equally spaced in frequency. We will denote this spacing by $\delta\omega$, and the resulting density $D$ for the discrete bath is then $D = 1/\delta\omega$. Since the evolution is unitary it is also reversible. The ability of the bath to induce apparently irreversible (thermodynamic) evolution in the system is due to the smallness of $\delta\omega$. We can think of the oscillators as dephasing over time, and we can expect that the resulting motion of the joint system will not repeat until all the oscillators have come back into phase. This happens on the timescale of $T = 1/\delta\omega = D$. Similarly, the thermodynamics that arises from large, many-body systems is possible only because the spacing between the energy levels of such systems is fantastically small. When $T$ is longer than all other timescales in the dynamics, including the timescale over which we wish to evolve the system, then we will refer to the frequencies of the oscillators as forming a \textit{quasi-continuum}. In order for the bath to induce rate equations for the system (to good approximation), we must choose the cut-off frequency $\Omega_{\ms{c}}$ to be larger than any frequency of the system. Together the requirements that $\delta\omega$ is small and $\Omega_{\ms{c}}$ is large imply that the total number of oscillators is large.


For our analysis, since the set of oscillators is discrete, we will label them with an integer index $k = 0, 1, \ldots, N$, so that there are $N+1$ oscillators, denote their respective annihilation operators by $a_k$ and their frequencies by $\omega_k$. The Hamiltonian of the bath is thus $H = \hbar \sum_{k=0}^N \omega_k a_k^\dagger a_k$, and for a uniform spectrum this is $H = \hbar \delta \omega \sum_{k=0}^N  k a_k^\dagger a_k$. We assume that the system to be coupled to the bath has discrete energy levels, and write the interaction Hamiltonian as  
\begin{equation} 
    H_{\ms{I}}  =  \hbar g X \sum_{k=0}^{N}  \left( a_k + a_k^\dagger \right) ,  
    \label{Hint1}
\end{equation}
As above, $X$ is a dimensionless Hermitian operator for the system, and its matrix elements are of order unity. Under the rotating-wave approximation, which requires that the decay rates generated by the interaction are much less than the non-degenerate transition frequencies, the interaction Hamiltonian becomes 
\begin{equation} 
    H_{\ms{I}}^{\ms{RWA}} =  \hbar g \sum_{k=0}^{N}  \left( A^\dagger a_k + A a_k^\dagger \right) ,  
    \label{Hint1b}
\end{equation}
in which $X = A + A^\dagger$, with $A$ being the upper triangular part of $X$ and $A^\dagger$ the lower triangular part. The diagonal elements of $X$ are taken to be zero, since their effect is only to induce shifts in the energy levels of the system.  

To analyze the structure of the coupling between the system and the oscillator bath it is useful to focus on a single system transition. If we denote the upper and lower levels of this transition by $|\mbox{e}\rangle$ and $|\mbox{g}\rangle$, respectively, and define the lowering operator $\sigma = |\mbox{g}\rangle \langle\mbox{e}|$, then the interaction between the system and bath that involves the transition is 
\begin{equation} 
    H_{\ms{int}}  =  \hbar g  \sum_{k=0}^{N}  \sigma \, a_k^\dagger + \sigma^\dagger \, a_k . 
    \label{Hint2}
\end{equation}
We now consider how this interaction couples the joint eigenstates of the system and the bath.  

\begin{figure}[tb]
\centering
\includegraphics[width=1\hsize]{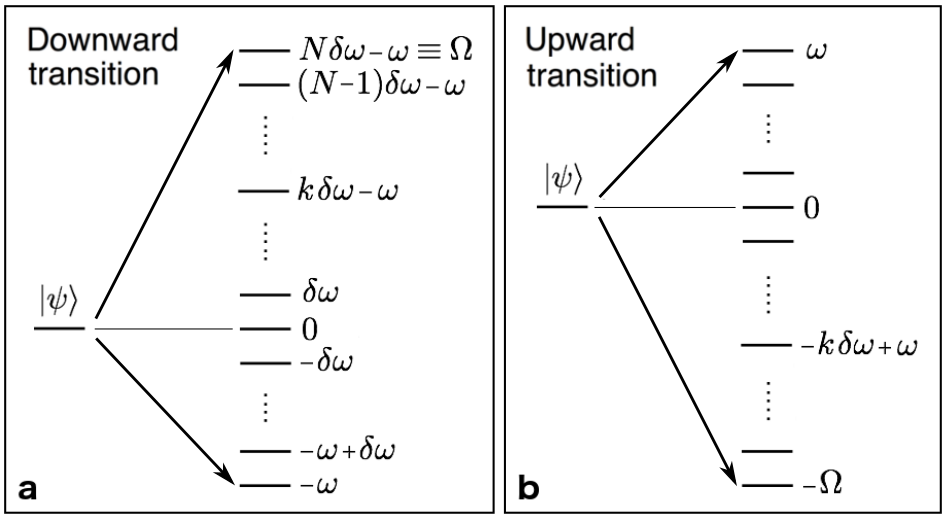}
\caption{A single transition (with frequency $\omega$) is coupled to a bath of harmonic oscillators, in which the frequencies of the oscillators are equally spaced in frequency from zero up to a ``cut-off'' frequency $\Omega_{\ms{c}}$. a) Here we show how the interaction couples the upper level of the transition to the lower level. In particular, when the bath is in any of its joint Fock states, and the system is in its upper level, then this initial state ($|\psi\rangle$ in the figure) is directly coupled to (and only to) a quasi-continuum of states, equally spaced in energy from $-\hbar\omega$ to $\hbar \Omega = \hbar (\Omega_{\ms{c}} - \omega)$. The matrix elements for these couplings are determined by the energy in each of the bath oscillators. b) The same as in (a), but this time the initial state of the system is the ground state. In this case the initial state is also directly coupled to a quasi-continuum, this time with the range  $-\hbar\Omega$ to $\hbar \omega$. For a given set of Fock states for the bath oscillators, the values of the matrix elements that couple the lower and upper levels are different.} 
\end{figure} 

\subsubsection{Zero temperature: coupling to a continuum }

Consider first what happens when the bath oscillators are at zero temperature (thus in their ground states) and the system starts in the upper level, $|\mbox{e}\rangle$. The coupling between the system and the bath allows the system to transition to its ground state $|\mbox{g}\rangle$ while simultaneously adding one quantum of energy to one bath oscillator. Thus the initial state is coupled to $N+1$ final states. Let us denote the joint vacuum state of the bath oscillators by $|\mathbf{0}\rangle$ and the state in which the $k^{\msi{th}}$ oscillator has 1 excitation (and the rest are in the vacuum state) as $|\mathbf{1}_k\rangle$.  With this notation the initial state is $|\psi\rangle = |\mbox{e}\rangle |\mathbf{0}\rangle$ and the $N+1$ final states are $|\phi_k \rangle = |\mbox{g}\rangle|\mathbf{1}_k\rangle$. Recall that we have chosen the frequencies of the resonators so that they are equally spaced in frequency with spacing $\delta\omega$. If we set the arbitrary reference point for energy so that the energy of the initial state $|\psi\rangle$ is zero, then the energies of the final states $|\phi_k \rangle$ are $E_k = \hbar (k \delta\omega - \omega)$ with $k = 0, \ldots, N$. That is, the final states form a uniformly dense spectrum covering the frequency range $[-\omega,\Omega]$ with 
\begin{equation}
     \Omega \equiv N\delta\omega-\omega = \Omega_{\ms{c}} - \omega.  
\end{equation}
This coupling structure is shown in Fig.~\ref{fig1}a. Thus the upper level of the transition is coupled to a quasi-continuum, which is precisely the scenario of Fermi's golden rule (FGR), with the twist that the upper end of the quasi-continuum is close to the bath cut-off frequency (assuming $\Omega_{\ms{c}} \gg \omega$) and the lower end is the (negative) frequency of the transition. 

Both the original derivation of Fermi's golden rule~\cite{Bransden00} and the Wigner-Weisskopf (WW) decay theory~\cite{Louisell} are essentially different approaches to obtaining an approximate solution to the above situation, namely the decay of a single transition at zero temperature (or equivalently a single level coupled to a quasi-continuum). Neither of these treatments provides much information about how the evolution of the system deviates from the dynamics of ideal rate-equations due to the fact that both $\Omega$ and $\omega$ are not infinite compared to the decay rate $\gamma$. It turns out that very much more information is revealed by later treatments of the FGR/WW scenario that provide a more complete solution, in particular those by Bixon and Jortner~\cite{Bixon68, Barnett03} and Fain~\cite{Fain88a, Fain88b} (there are also treatments by Davidson and Kozak~\cite{Kozak73a, Kozak73b, Kozak74} and Hillery~\cite{Hillery81}), and we exploit these in what follows.  

\subsubsection{Non-zero temperature I: a cascade \\ of couplings to continua}

Now consider what happens when the bath is at a temperature $T > 0$. Since the bath starts in a thermal state, we can consider it to be in a mixture of its energy eigenstates. As a result, the evolution of the joint system can be obtained by first determining this evolution for each of these eigenstates and then averaging over the evolutions. For every eigenstate of the bath each of the bath oscillators is in a ``number state'' (Fock state) in which it has a definite number of photons. Thus to analyze the dynamics it is sufficient to assume that each bath oscillator is in a number state, and that each of these number states is picked at random from the Boltzmann distribution for each oscillator at the bath temperature, $T$. 

\begin{figure}[tb]
\centering
\includegraphics[width=1\hsize]{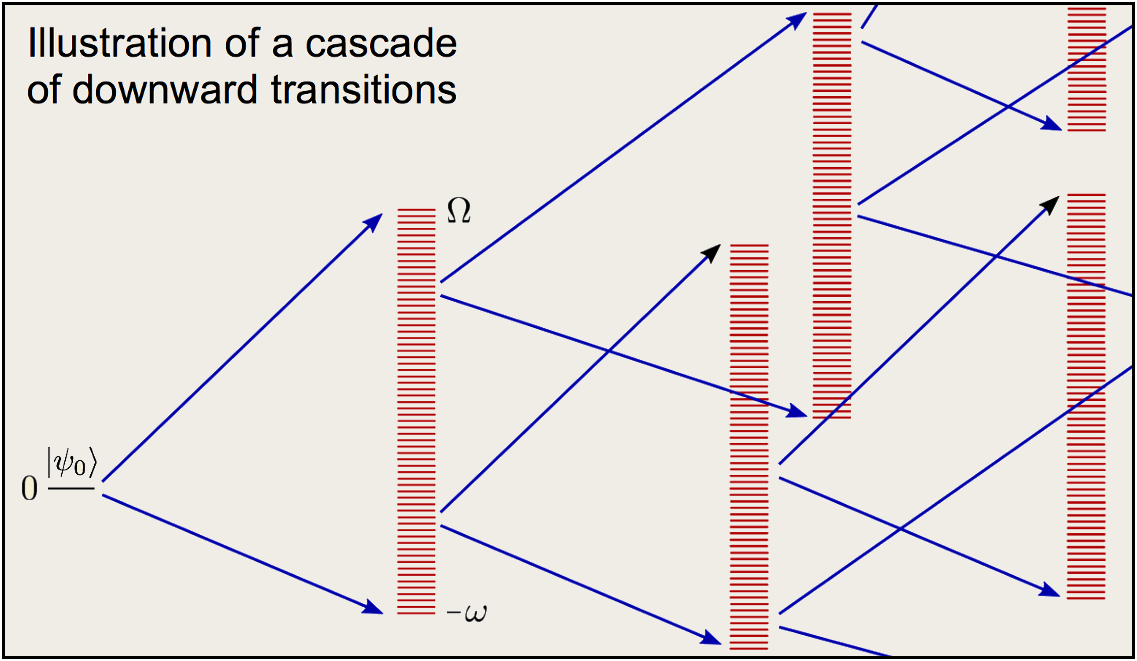}
\caption{Here we depict the coupling structure that results from an interaction with an oscillator bath under the rotating-wave approximation. Every initial eigenstate of the joint system is coupled to a quasi-continuum of joint eigenstates, and each state in the quasi-continuum is thus coupled to its own quasi-continuum, and so on, forming a ``cascade''. In the illustration every continuum coupling we depict corresponds to a downward transition, but most joint eigenstates are coupled to two separate continua, one mediating upward transitions and the other mediating downward transitions. An exception is those eigenstates corresponding to the ground state of the system, which is either coupled to a single continuum that mediates upward transitions (if the bath temperature $T$ is non-zero) or none at all if $T=0$. 
} 
\label{fig1}
\end{figure} 

Given that each oscillator starts in a number state, the situation is very similar to that discussed above for $T=0$. If the system starts in the upper level, $|\mbox{e}\rangle$, then in making a transition to the lower level, since any one of the oscillators can absorb a single photon, the change in energy of the joint system can take any of the values $E_k$ from the dense spectrum defined above with the range $[-\omega,\Omega]$. If the system starts in the lower level, $|\mbox{g}\rangle$, then because the oscillators are not (in general) in their ground states, they can emit a single excitation allowing the system to transition to the upper level $|\mbox{e}\rangle$. In doing so the total change in energy can take any value in the range $[\omega,-\Omega]$. Thus regardless of the temperature of the bath, every eigenstate of the system, if coupled at all, is coupled to a quasi-continuum with a range defined by the transition frequency and the bath cut-off frequency. 

So now imagine that the system starts in one of its energy eigenstates, and by virtue of its coupling to a quasi-continuum makes a transition to another of its eigenstates. Once in this second eigenstate the system is again coupled to a quasi-continuum, this time involving a entirely different set of states than the initial quasi-continuum, because the bath has also made a transition from its initial energy eigenstate to another. In fact, once the system has made its first transition the bath can be described as being in a superposition of all the states of the initial quasi-continuum. Every one of these states is coupled to it \textit{own} distinct quasi-continuum. Thus the oscillator bath consists of a ``cascade'' of couplings, each of which involves a single initial state coupled to a distinct quasi-continuum. This structure is represented pictorially in Fig.~\ref{fig1}. 

\subsubsection{Non-zero temperature II: randomized coupling}

We have now elucidated the coupling structure (the network topology) that connects the joint energy eigenstates of the system and bath. But we also need to know the \textit{values} of the matrix elements that couple the states together. The transition rates induced by the quasi-continuum coupling depend on these values. To determine these values we recall from the discussion above that, since the bath is in a thermal state, it is sufficient to assume that each of the bath oscillators is in a number state, in which the number of photons in each oscillator is picked from the Boltzmann distribution for that oscillator at temperature $T$. Without loss of generality we consider a downward transition for the system (in which one photon is added to the bath). Now, since the coupling Hamiltonian is given by Eq.(\ref{Hint2}), the matrix element that adds a photon to an oscillator that already contains $n$ photons is given by 
\begin{equation} 
    x(n) =  \hbar g \langle n+1 | a^\dagger | n \rangle  =  \hbar g \sqrt{n+1} .
\end{equation}
Next, for a typical energy eigenstate of the bath at non-zero temperature there will be a different number of photons in each oscillator (each being sampled independently from the Boltzmann distribution). The matrix elements that couple any given state to its quasi-continuum will vary randomly as one moves across the states of this continuum, since the matrix element for each state is sampled independently of the others. The distribution from which a given matrix element is sampled is the Boltzmann distribution for the photon number, $n$, which is a function of the frequency of the oscillator, and thus varies across the quasi-continuum. We know, for example, that for the oscillator with frequency $\omega$ the mean value of $n$ for the Boltzmann distribution is $n_T[\omega]$ (defined in Eq.(\ref{nT}) above), and thus the mean value of the square of the matrix element that couples to the oscillator with frequency $\omega$ (for a downward transition) is 
\begin{equation} 
    \langle [x(n)]^2 \rangle = \hbar^2 g^2 ( \langle n \rangle + 1 ) =  \hbar^2 g^2 (n_T[\omega]  + 1) . 
\end{equation} 

\subsubsection{Fermi's golden rule for a randomized coupling} 

The original golden rule was derived under the assumption that the matrix elements that couple the initial state to each of the states in the quasi-continuum are all equal. We found, however, in our analysis of the oscillator bath above, that the quasi-continuum couplings have matrix elements that are randomly sampled. We now demonstrate that a randomized continuum coupling also obeys Fermi's Golden rule, but with the square modulus of the (all identical) matrix elements replaced by the \textit{mean} of the square moduli of these matrix elements:  
\begin{equation}
    \gamma = \frac{2 \pi}{\hbar} \langle |x|^2 \rangle \mathcal{D}[E]      
    \label{genFGR} 
\end{equation} 
in which $\gamma$ is the decay rate, $x$ is the random variable that is sampled to obtain the independent, identically distributed matrix elements that couple the state to the quasi-continuum, and $\mathcal{D}[E]$ is the density of states (per unit energy) of the quasi-continuum. Equation (\ref{genFGR}) might be termed the ``stochastic'' version of the golden rule. 

\begin{figure}[tb]
\centering
\includegraphics[width=0.95\hsize]{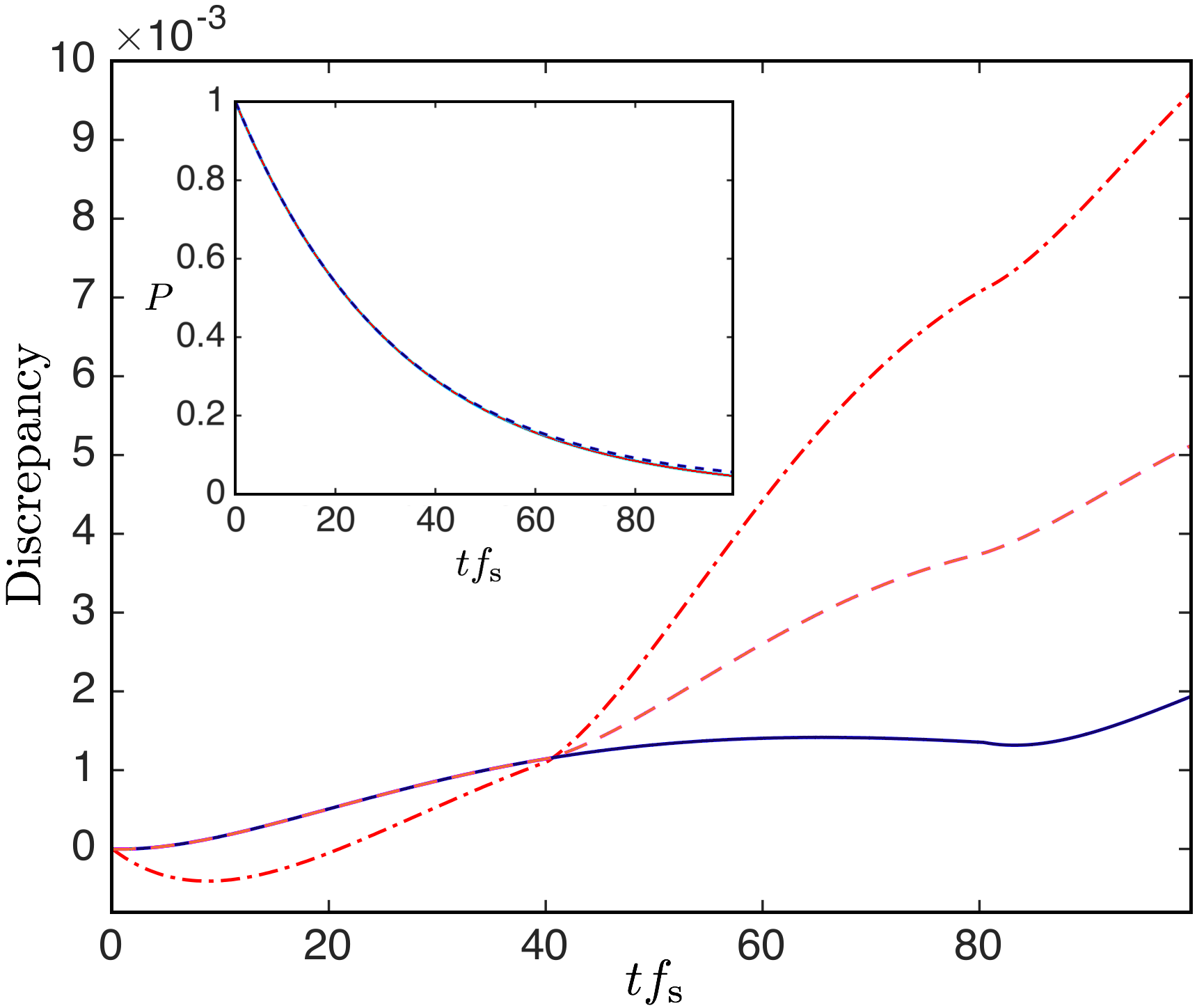}
\caption{Here we verify that Eq.(\ref{genFGR}), a modified version of Fermi's golden rule, holds when the matrix elements that couple the initial state to the continuum are independently sampled from a probability density. For our purposes the relevant density is the Boltzmann distribution, but we expect the same result to hold for any density that satisfies the central limit theorem. The inset compares the evolution due to Fermi's golden rule (solid line) in which the coupling is uniform across the spectrum to that resulting from a randomize coupling (dashed-line). In the main plot the three curves show how the discrepancy between the evolution due to the uniform and randomized couplings reduces as the number of levels in spectrum, $N$, is increased: $N=2^{15}$ (dot-dash); $N=2^{16}$ (dashed); $N=2^{17}$ (solid). 
} 
\label{fig3_randFGR}
\end{figure} 

We have not found a way to derive Eq.(\ref{genFGR}) analytically, so we resort to verifying it numerically. (Numerical studies performed for random matrix models also indicate the validity of Eq.(\ref{genFGR})~\cite{Gemmer09, Massimiliano03, Massimiliano03b, Lebowitz04, Gemmer06, Breuer06, Bartsch08, Silvestri14}.) To this end we define an arbitrary angular frequency unit $f_{\ms{s}}$, and simulate Fermi's golden rule with a symmetric spectrum of width $W = 40f_{\ms{s}} = 2\Omega$, $N = 2^{16} = 64\mbox{K}$ levels and a coupling rate $g = \sqrt{3} \mu$ with $\mu  = \sqrt{2}\times 10^{-3} f_{\ms{s}}$. These choices result in a damping rate of $\gamma = 2\pi g^2/N \approx 0.03$. We then perform the same simulation but this time with the matrix elements replaced by $\sqrt{n_{\ms{b}}} \mu$ in which $n_{\ms{b}}$ is sampled from the Boltzman distribution with mean $\langle n_{\ms{b}} \rangle = 3$. According to the ``stochastic golden rule'' this choice should give the same evolution as the first simulation for sufficiently large $N$. To compare the randomized coupling with the constant coupling we also average the evolution generated by the former over a thousand instances, so as to eliminate any residual fluctuations that come from the sampling as a result of the finite number of levels. We plot the respective evolutions for the two cases in the inset in Fig.~\ref{fig3_randFGR}. They are indeed close,  being barely distinguishable on the plot, although there is a small divergence discernible at later times. We examine this divergence more closely, and confirm that it reduces as the number of levels is increased, by plotting the difference between the two evolutions in Fig.~\ref{genFGR} for $N = 2^{15}, 2^{16}$, and $2^{17}$.   

Note that for the oscillator bath, for any given state, the probability distribution for the matrix elements that couple to the quasi-continuum varies across the continuum: this distribution is a function of the frequency of the oscillator that receives (or gives) a photon, because the Boltzmann distribution depends on frequency. Nevertheless, our analysis of the golden rule below shows that the width of the band of frequencies within the continuum that are important in generating the decay rate is on the order of this decay rate. Thus so long as the decay rate is small compared to the transition frequency (the weak-damping regime), we can replace the actual probability distribution with the Boltzmann distribution \textit{at the transition frequency}.   

\section{Obtaining the Master Equation from the Cascade Structure} 
\label{ObME}

We can now obtain the Markovian master equation directly from the structure we have elucidated above, combined with the generalization of Fermi's golden rule for a randomized coupling, Eq.(\ref{genFGR}), and a single additional conjecture. First, we see from the cascade structure that for every transition of the system, every joint eigenstate of the system and bath (these are states for which the system is either in the upper or lower state of the given transition, and each bath oscillator is in a Fock state) is coupled to its own continuum. Acting by itself, this coupling will induce dynamics that corresponds to a downward transition for the system with rate 
\begin{equation}
    \gamma_- = \frac{2 \pi}{\hbar} \langle (\hbar g\sqrt{n+1})^2 \rangle \mathcal{D}[E] =  \gamma (n_T[\omega]  + 1) 
\end{equation} 
and an upward transition with rate 
\begin{equation} 
    \gamma_+ = \frac{2 \pi}{\hbar} \langle (\hbar g\sqrt{n})^2 \rangle \mathcal{D}[E] =  \gamma n_T[\omega] ,  
\end{equation} 
Here we have defined the rate 
\begin{equation}
    \gamma \equiv 2 \pi g (\hbar g \mathcal{D}[E]) = 2 \pi g  \left( \frac{g}{\delta\omega} \right) , 
\end{equation} 
in which $\mathcal{D}[E]$ is the density of states of the quasi-continuum, and $\delta\omega$ is the frequency separation between the states in this continuum. 

Now, if the initial state of the joint system is a superposition of eigenstates, the continuum couplings   simultaneously induce dynamics for each eigenstate in the superposition. We now make our aforementioned ``additional conjecture'': we assume that the rate-equations (those induced by a set of distinct continuum couplings) for each of a set of states remain the same for each state when the system is in a superposition of those states. It follows immediately from this assumption that once the system has made a transition, so that the state of the joint system is now a superposition over the many states that make up one of the quasi-continua, the populations of each of these states now evolve under the rate equations induced by their coupling to their own quasi-continua, and so on down the cascade. Given this assumption it is clear why the rate equation behavior is not merely transient, but is maintained by the bath for all time. 

The rates derived above for the populations are precisely those of the Markovian master equation (MEW). The rate equations for the coherences then follow in a straightforward way, either by deriving them directly from the continuum coupling (similar to the derivation of Fermi's golden rule, see below), or by noting that they are the minimal rates of decoherence compatible with those for the populations.  This completes our ``derivation'' of the MEW from Fermi's golden rule. Note that the validity of the master equation can be considered evidence for the truth of the conjecture introduced above, since it is highly implausible that the dynamics could conspire to generate the master equation if it were not. 

\begin{figure}[tb]
\centering
\includegraphics[width=0.92\hsize]{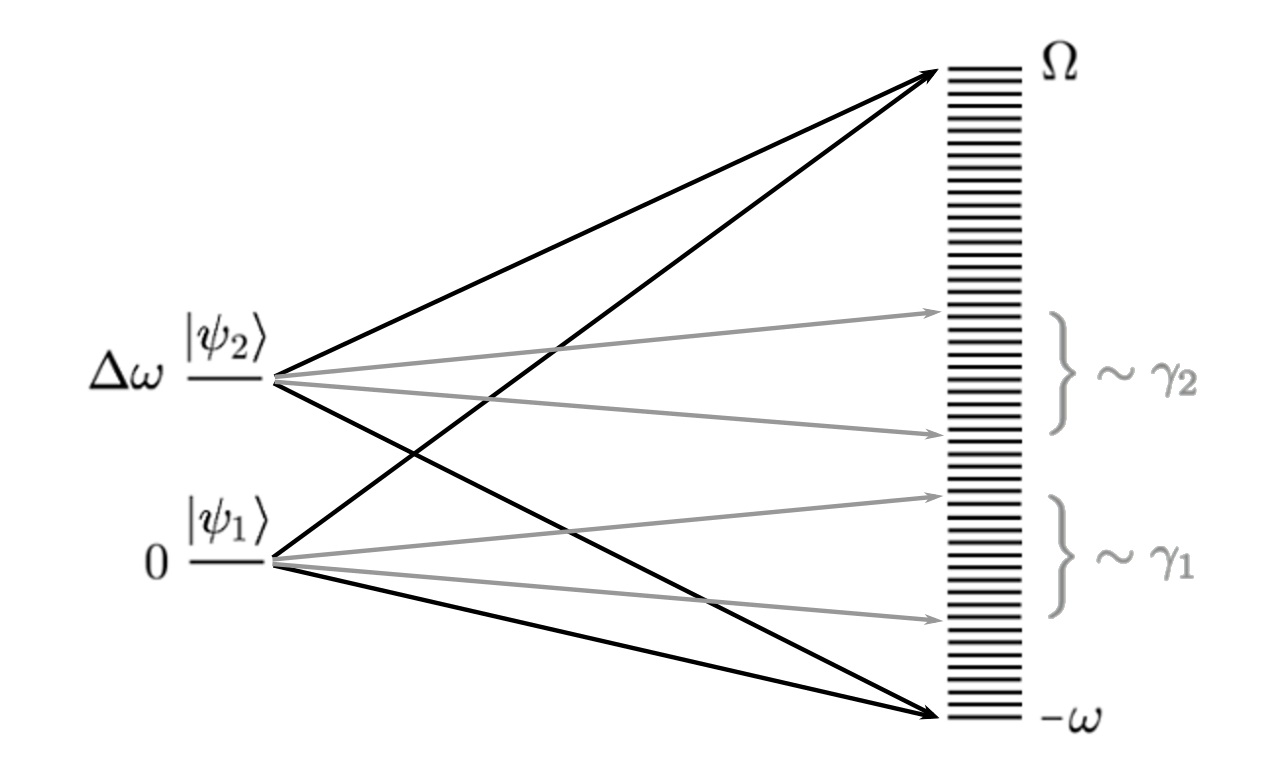}
\caption{A diagram showing why non-degenerate transitions must be well-separated ($\Delta\omega \gg [\gamma_1 + \gamma_2]/2$) in terms of the quasi-continuum coupling. All transitions are coupled to the same quasi-continua, and the effective width (in frequency) of this coupling is on the order of the decay rate. Since the transitions frequency determines \textit{where} the transition couples in the quasi-continuum, so long as the transition frequencies are far enough apart the two transitions are effectively coupled to separate continua, each of which induces a simple rate-equation. But if the frequencies are too close, then both levels are coupled to the same states in the quasi-continuum, which changes the overall behavior. In particular, the transitions are now effectively coupled to each other via the continuum, which induces a more complex dynamics.} 
\label{fig3}
\end{figure} 

\subsubsection{The requirement for well-separated transitions}  
\label{wellsep}

The cascade structure of the oscillator bath, combined with the golden rule, makes it intuitively clear how and why the rate equations of the MEW emerge from an interaction with the bath, and especially why these rate equations are sustained for all time. The cascade structure also helps us to understand why the MEW is only valid when the frequencies of the (non-degenerate) transitions of the system are separated by significantly more than their damping rates. In our analysis above we considered only a single transition, so now consider what happens if there are two, one with frequency $\omega$ and the other with frequency $\omega + \Delta \omega$. We note that since the system is coupled to a single quasi-continuum of harmonic oscillators, the downward (or, equivalently, upward) transition rates for both these transitions are induced by coupling to the \textit{same} quasi-continuum. That is, given an initial state for the system that is a superposition of the upper levels for the two transitions (and an initial eigenstate for the bath), the two initial states in the superposition are coupled to the \textit{same} quasi-continuum, and this is true of every quasi-continuum in the cascade. This situation is depicted in Fig.\ref{fig3}. 					

The only significant difference between the two transitions, as far as their interaction with the bath is concerned, is the location in the quasi-continuum with which they are resonant. If these locations are well-separated, we can expect that even though they are coupled to the same continuum neither will notice the presence of the other. But when they are close together we can expect a more complex joint dynamics to arise, mediated by the bath levels simultaneously coupled to both. The mutual ``interaction'' between the two transitions drops off with their energy difference because the levels in the quasi-continuum have less effect on a given transition as they move out of resonance with it. By analyzing the dynamics of the continuum coupling (which we do in the following section), one can see that the interval of the quasi-continuum with which a given transition interacts is on the order of the decay rate $\gamma$. Thus the condition $\Delta \omega \gg \gamma$ can be expected to preserve the dynamics of the master equation. Outside that regime the dynamics of any two nearby transitions is significantly more complex because they are effectively coupled together. We discuss this further in Section~\ref{3dev}.

\subsubsection{The origin of the thermal steady state}

Finally, it is interesting to note how the oscillator bath ensures the thermal (Boltzmann) steady-state. If we have a discrete set of states for which the dynamics is described by transition rates between pairs of states, then it can be shown that the ratio of the populations of states A and B in the steady state is given by $P_{\ms{A}}/P_{\ms{B}} = \gamma_{\ms{B}\rightarrow{\ms{A}}}/\gamma_{\ms{A}\rightarrow{\ms{B}}}$, in which $\gamma_{\ms{B}\rightarrow{\ms{A}}}$ is the transition rate from B to A and vice versa, so long as the full set of pairwise transition rates gives a consistent set of populations~\cite{Alicki76}. A steady-state that satisfies this relationship is said to arise from \textit{detailed balance} (balancing of each transition). 
 
The master equation will achieve the Boltzmann steady-state via detailed balance if the pairs of upward and downward transition rates satisfy   
\begin{equation}
    \frac{\gamma_+}{\gamma_-} = \exp\left[ - \hbar\omega/(k_{\ms{B}} T) \right],  
\end{equation} 
in which $\gamma_+$ and $\gamma_-$ are upward and downward rates for a transition with energy gap  $\hbar\omega$, $k_{\ms{B}}$ is Boltzmann's constant, $\omega$ is the transition frequency, and $T$ is the bath temperature. Since the densities of all the quasi-continua are the same, and independent of the bath temperature, the only way that the upward transition rates can be different from the downward rates, required by the above condition, is if the matrix elements that couple the lower state to its quasi-continuum are different from those for the upper state. That this is true is due to the fact that the coupling to the bath involves the mode operators $a$ and $a^\dagger$, and these give different  matrix elements for subtracting ($a$) and adding ($a^\dagger$) a photon to each oscillator. These matrix elements, combined with the condition that all the oscillators are in their respective Boltzmann states, conspire to ensure that the rates, when averaged over the thermal states, satisfy the above condition for thermalization. Naively one would expect that only the thermal state of the oscillators would be required to thermalize the system. We will not investigate this question further here, but the need for a specific interaction would appear to be a special, rather than generic, property of the oscillator bath model.  

\section{Solving the Hamiltonian of Fermi's golden rule}  
\label{secFGR}

In the previous section we saw that the oscillator bath consists of a cascade of coupling structures, each of which corresponds to that of Fermi's golden rule (FGR). We now examine the analytic solution to the golden rule because it provides insight both into the emergence of rate equations and the way in which the dynamics deviates from these equations as the required conditions are relaxed. We note first that since quantum evolution is unitary, it must always consist of a sum of oscillations at a set of eigenfrequencies. The weightings in this sum are determined both by the energy eigenstates and the initial state. On the contrary, the evolution of the population of a state under a rate equation does not oscillate, but instead undergoes exponential decay to a steady-state. Thus in order for a rate equation to emerge from the unitary evolution of quantum mechanics, the eigenvalues and eigenvectors of the evolution must conspire in a special way: the resulting sum of complex exponentials, oscillating at the eigenfrequencies, must reproduce a decaying exponential. A coupling to a uniformly dense spectrum is, as far as we are aware, the only situation in which this happens. As we will see it is only in the limit in which the dense spectrum is very dense, and when thus spectrum continues to positive and negative infinity, that an exact exponential decay is produced. 


We consider a state $|\psi\rangle$ that will be initially populated, and assign to it an energy of $0$. We then couple it to a set of $N+1$ states that we denote by $|k\rangle$ and that we will call the ``final states''. We take this coupling to have the following form: i) each final state is coupled to the initial state by a matrix element with the same value, $i \hbar g$; ii) the final states are equally spaced in energy, and the gap between two adjacent states is $\hbar \delta\omega$; iii) the energy range for the set of final states will be taken to be either symmetric and equal to $\hbar[-\Omega,\Omega]$ (in which case $\Omega \rightarrow \infty$ is the Brixon-Jortner model~\cite{Barnett03, Bixon68}) or asymmetric and equal to $\hbar[-\omega,\Omega]$ with $\omega < \Omega$. 

The scenario described above can be solved (almost) exactly with little more effort that that required for the perturbative treatment usually found in textbooks for Fermi's golden rule. While the following analysis is essentially that of Bixon and Jortner~\cite{Bixon68, Barnett03}, with the benefit of an extension obtained by Fain~\cite{Fain88a, Fain88b}, we feel it is worth presenting it here in a way that is tailored to our purposes. We begin by writing the state of our system, which consists of the initial state $|\psi\rangle$ and the states $|k\rangle$ that form the quasi-continuum, as    
\begin{align}
   |\phi \rangle = d |\psi\rangle + c_k  |k\rangle . 
\end{align} 
The equations of motion are then 
\begin{align}
   \dot{d} & =   g \sum_k c_k  ,   \label{ee1}  \\
   \dot{c_k} & =  k \delta\omega c_k - g c    , \;\;\; \forall k       \label{ee2}  
\end{align}
where $g$ is the value of the coupling to each of the states $|k\rangle$ and $\delta\omega$ is the spacing between the oscillator frequencies. Here we have labelled the final states using  
\begin{align}
   k = -K, \ldots , -1, 0, 1, \ldots , M , 
   \label{krange}
\end{align}
so that the low end of the spectrum is given by $-\omega = -K \delta\omega $ and the high end by $\Omega = M\delta\omega$. If the spectrum is symmetric so that $\omega = \Omega$ then we take $N$ to be even so that $K = M = N/2$. 

We want to find the evolution of $d$. To begin with we will take the quasi-continuum to be symmetric so that $K = M$. If we write the quantum state as the vector $\mathbf{v} = (d,c_{-K},\ldots,c_{M})$ and the equations of motion as $\dot{\mathbf{v}} = A \mathbf{v}$, then the solution is given by the eigenvectors and eigenvalues of $A$. If we denote the eigenvalues and eigenvectors of $A$ by $\lambda_n$ and $|v_n\rangle$, respectively, then the evolution of $c$ is given by 
\begin{equation}
   d(t) = \sum_{n=0}^{N+1}  |\langle \psi | v_n \rangle|^2  e^{-i\lambda_n t}  = \sum_{n=0}^{N+1}  |v_{n0}|^2  e^{-i\lambda_n t}  ,  
   \label{eq2}
\end{equation} 
in which $v_{n0}$ is the first element of the $n^{\msi{th}}$ eigenvector of $A$. If the eigenvalues of $A$ are approximately equally spaced, and we will find that they are, then the population of the initial state, $P = |d(t)|^2$, will decay exponentially only if the expression above is close to the Fourier series for this exponential. This in turn requires that the amplitudes of the first elements of the eigenvectors, $v_{n0}$, must be a Lorentzian function of the eigenvalues, $\lambda_n$. Luckily one can obtain an analytic expression for $v_{n0}$ for large $N$. 

To determine the coefficients $v_{n0}$ we start, naturally, with the equations for the eigenvalues and eigenvectors of $A$, namely $A |v_n\rangle = \lambda_n |v_n\rangle$. In terms of the elements of the eigenvectors these equations are 
\begin{align}
g  \biggl( \sum_k v_{nk}  \biggr) & = \lambda_n v_{n0} ,   \label{eveq1} \\
g v_{n0} + \delta\omega [k - M - 1] v_{nk} & = \lambda_n v_{nk} ,   \label{eveq2}
\end{align}
where $k = 1, \ldots, N+1$ and $v_{nk}$ is the $k^{\msi{th}}$ element of $|v_n\rangle$. We can easily rearrange these equations to write the coefficients $v_{nk}$ in terms of $v_{n0}$ as     
\begin{align}
   v_{nk} = \left( \frac{g}{\lambda_k - [k - M - 1]\delta\omega} \right) v_{n0}, 
\end{align} 
Now using this expression and the normalization condition for the eigenvectors $|v_{n0}|^2 + \sum_k |v_{nk}|^2 = 1$, we obtain the first element of $|v_n\rangle$ as a function only of the eigenvalue $\lambda_n$: 
\begin{align} 
   |v_{n0}|^2 = \left( 1 + \sum_{j=-M}^M \frac{g^2}{(\lambda_n - j\delta\omega)^2} \right)^{-1} . \label{dlwt}
\end{align} 
This expression does not look like a Lorentzian function of $\lambda_n$, but some tricks with infinite series reveal that it is. We first use the relation  
\begin{align} 
  \sum_{j=-\infty}^{\infty} (x - j q)^{-2} =  (\pi/q)\mbox{cosec}^2(\pi x/q)  
\end{align} 
to replace the sum in Eq.(\ref{dlwt}), which is an arbitrarily good approximation as $N \rightarrow \infty$. 

We now return to the equations for the eigenvalues of $A$, Eqs.(\ref{eveq1}) and (\ref{eveq2}), where eliminating the elements of the eigenvectors allows us to write an equation for the eigenvalues alone, which is  
\begin{align}
\frac{\lambda}{g^2} = & \sum_{j=-M}^{M}(\lambda-j\delta\omega)^{-1} = \frac{1}{\lambda} + 2\lambda^2 \sum_{j = 1}^{M}\frac{1}{\lambda^2 - [j\delta\omega]^2} \nonumber \\ 
    \rightarrow & \; (\pi/\delta\omega)\cot(\pi\lambda/\delta\omega) \;\, \mbox{as} \;\, M\rightarrow\infty.
\end{align} 
With this equation for $\lambda$, along with the trigonometric relation $\cot^2\theta = \mbox{cosec}^2\theta - 1$ we can replace the cosec function in the expression for $|v_{n0}|^2$ to obtain 
\begin{align}
   |v_{n0}|^2 = \frac{g^2}{\left(\frac{\pi g^2}{\delta\omega}\right)^{\! 2} \left[ 1 + \left(\frac{\delta\omega}{g\pi}\right)^2 \right]  + \lambda_n^2 } . 
\end{align} 
This is indeed a Lorentzian function of the eigenvalues, and is the reason that an interaction with a quasi-continuum, and thus a bath of oscillators, generates rate equations.  

We now show that the eigenvalues $\lambda_j$ come in positive/negative pairs and are equally spaced, so that the expression for $c(t)$ is indeed a close approximation to the Fourier series for a decaying exponential. Both these facts result from the equation we obtained above for the eigenvalues for large $N$, namely, 
\begin{align} 
   \lambda \left(\frac{\delta\omega}{\pi g^2}\right)  = \cot \left(  \lambda \left[ \frac{\pi}{\delta\omega} \right]\right) . 
   \label{lam}
\end{align} 
This relation gives the eigenvalues as the intersections of a straight line (the LHS) with a cotangent function that has the period $\delta\omega$ (the RHS). Adjacent eigenvalues will have a separation close to $\delta\omega$ if the straight line has a slope that is sufficiently gentle, and this is the case when 
\begin{align} 
  \left( \frac{\delta\omega}{g} \right)^2 \ll 1 .   \label{c1} 
\end{align} 
The eigenvalues come in positive/negative pairs because Eq.(\ref{lam}) is symmetric in $\lambda$.  

We can now examine the expression for $d(t)$ and extract the decay rate, $\gamma$. We note first that the coefficients of the Fourier series, $|v_{n0}|^2$, while Lorentzian, are not normalized correctly unless $(\delta\omega/g)^2 \ll \pi^2$. However this condition is already subsumed by the condition obtained above for equal spacing of the eigenvalues. Given the condition in Eq.(\ref{c1}) we can write 
\begin{align}
   d(t) & \approx \sum_{\substack{n=-(M+1) \\ n\not=0}}^{M+1}  \left[ \frac{g^2}{\left(\pi g^2/\delta\omega\right)^{\! 2}  + \lambda_n^2 } \right]  e^{-i\lambda_n t} \nonumber \\
      & =  \sum_{n=1}^{M+1}  \left[ \frac{ 2 g^2}{  \gamma_d^2 + \lambda_n^2 } \right] \cos(\lambda_n t) ,   \;\;\; \lambda_n \approx \left( n - \frac{1}{2} \right) \delta\omega , 
   \label{eqa4}
\end{align} 
where we have defined $\gamma_d \equiv \pi g^2/\delta\omega$. This expression for $d(t)$ is the key result as far as Fermi's golden rule is concerned; it shows that the unitary quantum dynamics of a coupling to a continuum results in a rate equation for the amplitude (and thus the population) of the initial state. Given the form of the Fourier transform of a decaying exponential, we can now read off the decay rate of $d(t)$ which is $\gamma_d$. The decay rate of the population $P = |d|^2$ is then $\gamma = 2 \gamma_d$, and this is 
\begin{equation}
   \gamma = 2\pi g \left( \frac{g}{\delta\omega} \right) = \frac{2\pi}{\hbar} (\hbar g)^2  \mathcal{D}[E],   \label{FGR}  
\end{equation} 
where $\mathcal{D}[E] = 1/(\hbar \delta\omega)$ is the density of states per unit energy of the quasi-continuum. This is Fermi's golden rule. 

In our analysis so far we have taken the quasi-continuum to have the symmetric spectral range $[-\Omega,\Omega]$, but we know that the oscillator bath contains quasi-continua with the asymmetric range $[-\omega,\Omega]$. We note that since the coefficients of the Fourier series in Eq.(\ref{eqa4}) are given by a Lorrentzian with width $\gamma$, the part of the spectrum closest to the energy of the initial state contributes most to the evolution, and parts of the spectrum that are much further away than $\gamma$ contribute very little. Thus as long as both $\omega$ and $\Omega$ are sufficiently large the asymmetry of the spectrum can be safely ignored. However, if we want to know how the finite value of $\omega/\gamma$ affects the dynamics, and distinguish this from the effect of a finite cut-off frequency $\Omega$, then we need to include the fact that the spectrum is asymmetric. In the next section we show that this can be done quite easily. 

We can extract from the above analysis two further conditions that are required for the emergence of rate equations. The first is that the inverse of the separation between the levels in the quasi-continuum, $T_{\delta} \equiv 2\pi/\delta\omega$ (being the density of the spectrum in units of time), must be much greater than any other timescale of interest, and in particular that of the damping time $T = 1/\gamma$ (which is usually the next largest timescale in the problem). The resulting condition is $g \gg \delta \omega/ (2 \pi)$.  The final condition is that the Markovian timescale should be much smaller than the timescale of the dynamics in which we are interested. The smallest timescales in which we can be interested as far as the system is concerned is that of the system's transitions. Characterizing the transition frequencies by $\omega \gg \gamma$, the resulting condition is  
\begin{equation} 
   \sqrt{N} =  \left( \frac{\Omega}{\delta\omega}  \right)^{\! 1/2}   \gg   \left( \frac{\omega}{\delta\omega} \right)^{\! 1/2}    \gg  \frac{g}{\delta\omega}  \gg 1 . 
\end{equation}
These relations encapsulate all but one of the conditions for the emergence of effective rate equations from a coupling to a continuum (the exception is the condition that non-degenerate transitions be well-separated), and confirm that the number of states in the dense spectrum must be very large. These conditions have been obtained via various methods in previous derivations of the master equation~\cite{Breuer07, Alicki89, Davies1974, Redfield57, Redfield65}.

\section{Three distinct ways in which an open system deviates from the master equation} 
\label{3dev}

For real many-body thermal baths the density of levels in the near continuum is very, very high (think Avogardo's number), so we can assume that we will not see deviations from rate-equations due to the finite density of the quasi-continuum. From our analysis above we can identify three distinct ways in which the bath or system parameters may deviate from the ``ideal'' dynamics of the master equation: i) the value of the cut-off $\Omega$ may not be sufficiently large (compared to $\omega$ and $\gamma$); ii) the system frequency $\omega$ may not be sufficiently large (compared to $\gamma$); iii) the difference in frequency, $\Delta \omega$, between two transitions may not be sufficiently large (compared to $\gamma$). We now use our analysis above to determine the distinct ways in which each of these situations changes the dynamics. 

\subsubsection{The effect of a finite bath cut-off frequency: the Markovian timescale $\tau$} 

The expression for the evolution of the amplitude of the initial state, $d$, is a cosine series. Since the derivative of cosine at 0 is zero, so is the derivative of $d$ at $t=0$. The initial evolution of $d$ is thus \textit{second order} in time, which is demanded by the unitary evolution of quantum mechanics and the fact that the bath and system are initial in a separable state. The derivative of an exponential decay at $t=0$ is equal to (minus) the decay rate multiplied by the initial value of $d$. Thus the cosine series cannot match exponential decay at the initial time. There is a time $\tau$ that much elapse before the rate of change of the cosine series has decreased by enough to reach the value $-\gamma$. It is only after this time has elapsed that the evolution of the populations is given by rate equations. This time is called the ``Markovian timescale'' (or equivalently the ``non-Markovian timescale'') and it is only after this time that the evolution of the system is governed by an effectively Markovian master equation. The result of a nonzero value of $\tau$ is that the actual evolution of the populations due to the bath lags that given by the MEW by approximately $\tau$. Because of this the effect of a non-zero value of $\tau$ has been referred to by some authors as the ``initial slip''~\cite{Gnutzmann96}. So long as this lag is much smaller than the decay rate(s) themselves, the error in the values of the populations is very small. 

We can use the cosine series to obtain an approximate value for $\tau$. First it is clear than $\tau$ is limited by the highest frequency in cosine sum, since $d(t)$ will not change on a timescale that is smaller than the inverse of this frequency. Recall that the derivative of $d(t)$ is initially $0$, and must decrease until it reaches the value $-\gamma$, Similarly the curvature, which is initially negative, must increase to reach the value $\gamma^2$. 
Differentiating Eq.(\ref{eqa4}) gives us 
\begin{equation}
   d''(t) \approx - \frac{\gamma \,\delta\omega}{\pi} \sum_{n=1}^{M+1}  \left[ \frac{ \lambda_n^2}{  \gamma_d^2 + \lambda_n^2 } \right] \cos(\lambda_n t) 
\end{equation}
The fact that $\Omega \gg \gamma$ means that for most values of $n$ in the above sum the expression in square brackets is unity, so that 
\begin{equation}
   d''(t) \approx - \frac{\gamma \,\delta\omega}{\pi} \sum_{n=1}^{M+1} \cos[(n-1/2)\delta\omega t] 
\end{equation}
and $d''(0) \approx -\gamma \Omega/\pi$. All the cosines are initially unity, and as $t$ is increased they all start to decrease. The ones with the highest frequency (the largest values of $n$) are the ones that cross zero first to go negative, and as $t$ continues to increase, the value of $n$ for which $\cos[(n-1/2)\delta\omega t]$ becomes negative sweeps down to smaller $n$. Thus in order for $d''(t)$ to increase to zero, approximately half of the cosines in the sum must go negative, meaning that $\cos[(M/2-1/2)\delta\omega t]$ hits zero.  The Markovian timescale $\tau$ is therefore given approximately by $(M/2-1/2)\delta\omega \tau \approx (\Omega/2) \tau = \pi$ or 
\begin{align}
   \tau \approx 2 \pi /\Omega .  
\end{align} 
Previous derivations and discussions of the Markovian timescale can be found in, e.g.~\cite{Alicki89, Hu92, Breuer07, Davies1974, Redfield57, Redfield65}. 

\begin{figure}[tb]
\centering
\includegraphics[width=1\hsize]{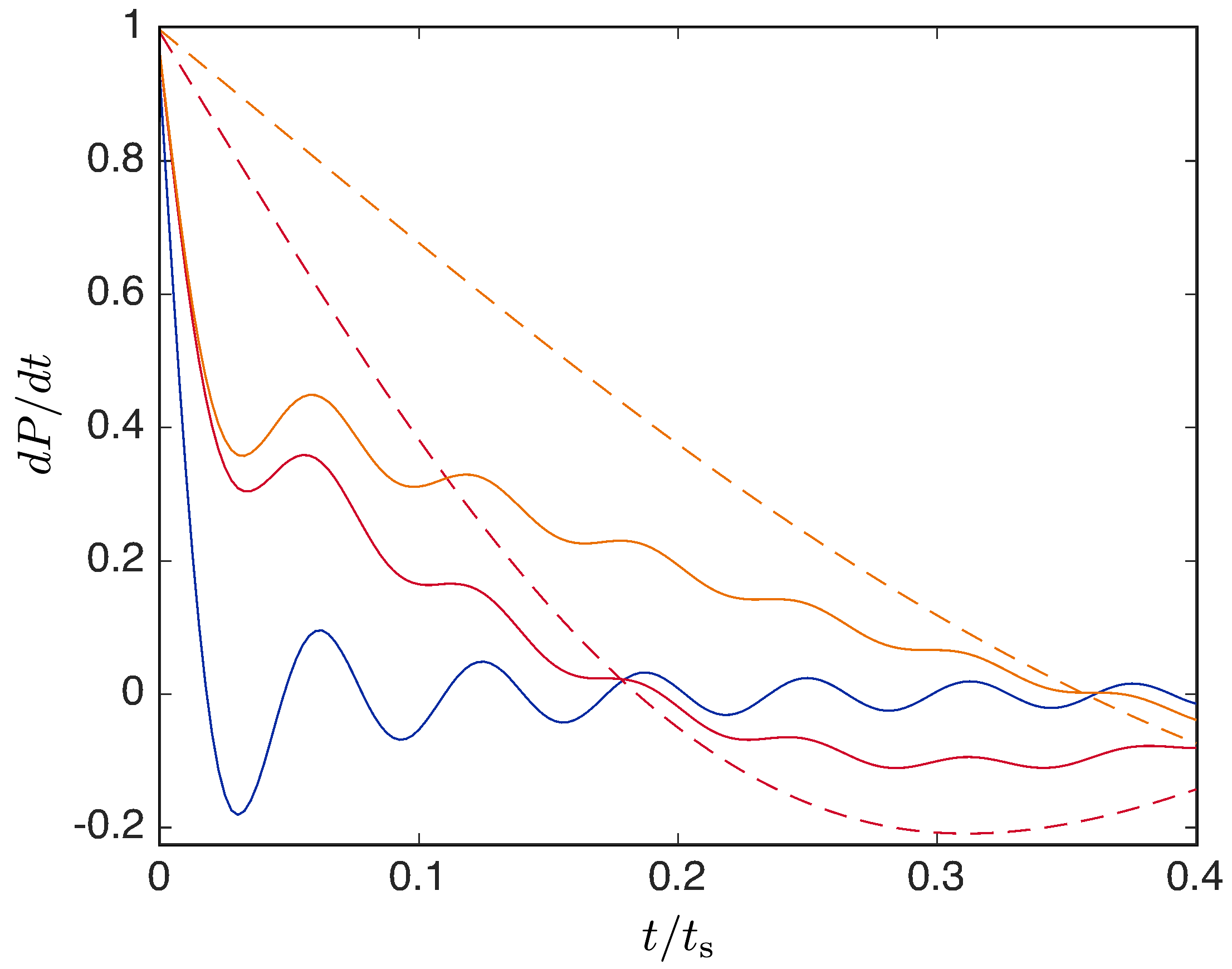}
\caption{Here we compare the deviation from the master equation incurred by a finite value of $\Omega/\gamma$ to that due to a finite value of $\omega/\gamma$ (given that $\omega \ll \Omega$). The latter results in a quasi-continuum with an appreciably asymmetric spectrum. Here we have set $\gamma = 0.2$ (in arbitrary units) and we plot the deviation between the derivative of the population of the initial state, $P = d^2$, to that for an ideal exponential decay. The blue curve shows the deviation when $\Omega = \omega = 100$. The red curves compare the effect of reducing $\omega$ by a factor of 20 (solid line) to that of reducing both $\Omega$ and $\omega$ by the same factor (dashed line). Similarly the orange curves compare the effect of reducing $\omega$ by a factor of 50 (solid line) to that of reducing both $\Omega$ and $\omega$ by this factor (dashed line).} 
\label{fig4}
\end{figure}

\subsubsection{The effect of a finite transition frequency $\omega$: distortion due to an asymmetric spectrum}

To determine the physical effects of the fact that $\Omega/\omega$ and $\omega/\gamma$ are finite, we need to know what happens when the spectrum is asymmetric, meaning that the transition energy $\hbar\omega$ is not in the center of the spectrum. It is now that the results of Fain's analysis of a coupling to a continuum is invaluable~\cite{Fain88a, Fain88b}. We have seen that for a symmetric spectrum the eigenvalues, and thus the Fourier series, has a range that is very nearly equal to that of the dense spectrum to which the initial state is coupled. Determining the behavior when the spectrum is asymmetric is now made simple by using the fact that the spectrum is \textit{very dense}. Fain analyzed this situation by first taking the limit in which the dense spectrum becomes a continuum. The resulting expressions reveal that the energy of each resulting eigenstate remains very close to that of a state in the dense spectrum, so that the range of the eigenstate energies matches closely that of the initial spectrum. We can therefore obtain a good approximation to the evolution simply by replacing the upper and lower limits in the Fourier series, Eq.(\ref{eqa4}), with those of the actual spectrum to which the transition is coupled. This means that the upper end of Fourier series is still $\Omega$, but the lower end becomes $-\omega$ (given that we have set the energy of the initial state to be zero). The resulting ``lopsided'' Fourier series can be decomposed into the original cosine series in Eq.(\ref{eqa4}) along with a sine series over the frequency range $[\omega,\Omega]$. Writing the resulting series as integrals we have 
\begin{align}
  d(t) \approx \int_{0}^{\Omega} \!\! F(\lambda) \cos(\lambda t) d\lambda   +  i \int_{\omega}^{\Omega} \!\! F(\lambda) \sin(\lambda t) d\lambda, 
\end{align} 
in which $F(\lambda) = (\gamma/\pi) /[\gamma_d^2 + \lambda^2]$. 

The fact that the distortion of the exponential decay is given by a sine series means that it has little effect at $t=0$ and thus does not have a major impact on the Markovian timescale. It does nevertheless cause a distortion of the exponential decay, a distortion whose oscillations are strongest in the vicinity of the transition frequency, since the function $F$ is largest at this frequency. The overall size of the distortion can be obtained from the overall size of the coefficients of the sin series as a proportion of the coefficients of the cosine series. We therefore integrate the Lorentzian over the range $[\omega,\Omega]$, using the fact that $\omega \gg \gamma$, and divide this by the integral over the full range $[0,\Omega]$. This calculation gives the size of the distortion as $\gamma/(2\pi\omega)$. The error due to the finite transition frequency, $\omega$, thus scales as $1/\omega$.  

In Fig.\ref{fig4} we compare the deviation from exponential decay (specifically, the deviation in the derivative of the population $P=d^2$ from that of an exponential decay) induced by a finite value of the cut-off $\Omega$ to that induced by a finite value of $\omega$. For these simulations we chose $\gamma = 0.2/t_{\ms{s}}$ where $t_{\ms{s}}$ is an arbitrary timescale. The blue curve is the error when $\Omega  = \omega = 100/t_{\ms{s}}$ so that the spectrum is symmetric. We then compare the effect of reducing \textit{both} $\omega$ and $\Omega$ (so that the spectrum remains symmetric and the deviation is due purely to the non-zero value of the Markovian timescale) to that of reducing only the transition frequency $\omega$ so that the spectrum become asymmetric. We see from this that when we reduce both $\omega$ and $\Omega$ the Markovian timescale simply increases so that the entire error curve is merely stretched on the time axis. When we change only $\omega$ we see that initially the error still decreases almost as rapidly as that for the blue curve ($\Omega  = \omega = 100/t_{\ms{s}}$), and while the error no longer crosses zero on the timescale $\tau$, the finite value it reaches on this timescale is not far from that due to the oscillations of the error that occur for the blue curve. This provides a justification for the traditional view that the Markovian timescale is due to the finite value of the cut-off $\Omega$ rather than $\omega$, although clearly the situation is nuanced and is to some extent a matter of definition. 

\begin{figure*}[tb]
\centering
\includegraphics[width=1\hsize]{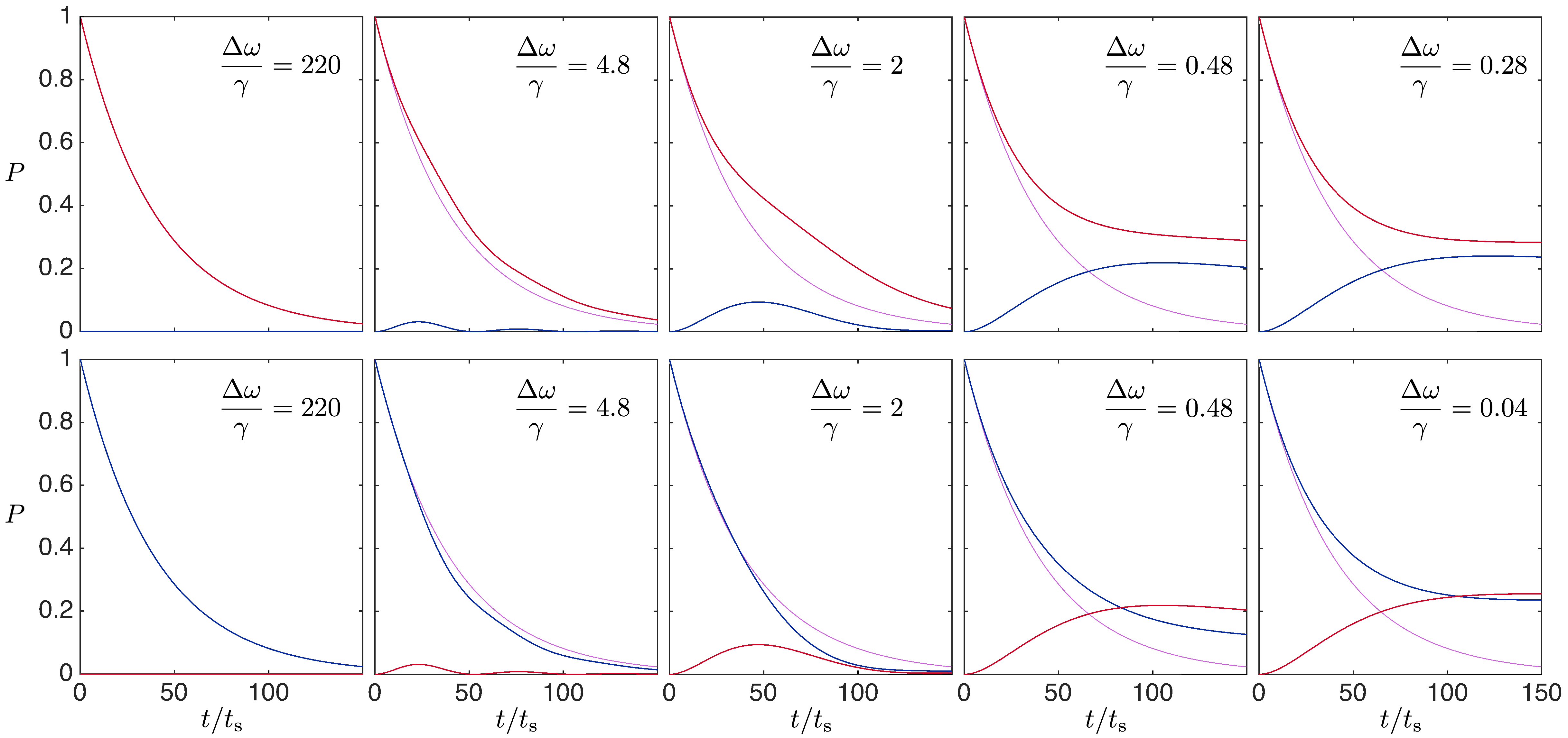}
\caption{The decay of two non-degenerate transitions (at zero temperature) whose behavior is outside that of the standard Markovian master equation due to their proximity. The upper set of plots shows the evolution when the transition with the higher frequency is the only one populated, and the lower set shows the evolution when only the lower transition is populated. For both plots the population of the higher-frequency transition is shown in red and that of the lower frequency transition in blue. The mauve curve is the usual exponential decay that occurs when the two levels are well separated. The frequency separation of the two transitions is denoted by $\Delta \omega$, the decay rate of the two levels by $\gamma$, and $t_{\ms{s}}$ is an arbitrary timescale. For these simulations the frequency of the lower transition is fixed at $\omega = 5.5/t_{\ms{s}}$, the decay rate of both levels is $\gamma = 0.025/t_{\ms{s}}$, the cut-off frequency of the bath is $\Omega = 32/t_{\ms{s}}$, and we used a bath with 128,000 oscillators (each oscillator contributes one state to the quasi-continuum). 
} 
\label{fig5}
\end{figure*}

\subsubsection{The effect of an asymmetric spectrum II: the Lamb shift}
\label{shift}

The Lamb shift is the well-known effect in which an interaction with an oscillator bath (and more generally the interaction of a localized system with a field) causes an effective shift in the transition frequencies of the system. This shift is especially curious in that it depends on the cut-off frequency $\Omega$, and in particular it is only finite so long as $\Omega$ is finite. This implies that the coupling of a localized system to a field must always drop off in some way at high frequencies. The Lamb shift also depends on exactly \textit{how} the coupling decreases with frequency. In our simple bath model we have chosen the coupling to be constant for all frequencies up to $\Omega$ and zero for all beyond. This is referred to as a ``hard cut-off'' at $\Omega$. 

When the spectrum of the quasi-continuum is symmetric (that is, the energy of the initial state is at the center of the quasi-continuum) there is no Lamb shift. One can view the Lamb shift as arising from the fact that for an oscillator bath the spectrum is asymmetric. It comes from the fact, which we ignored in the previous section, that the discrete frequencies that make up the approximate lopsided Fourier series for $d(t)$ are not symmetrically placed about zero, and thus give to $d(t)$ an overall frequency shift superimposed on top of the decaying exponential. This effectively shifts the frequency of the initial state (corresponding to the upper state in the transition) with respect to the lower state. For a quasi-continuum whose frequency range is $[-\Omega_-,\Omega_+]$, and for which the density of states per unit frequency is constant, one can calculate the following approximate expression for the Lamb shift, $\Delta \omega_{\ms{L}}$ using contour integration methods~\cite{Louisell, Barnett03}: 
\begin{equation}
      \Delta \omega_{\ms{L}} = \left( \frac{\gamma}{2\pi} \right) \ln \left[ \frac{\Omega_+}{ \Omega_-} \right] + \mathcal{O} \left( \frac{\gamma^2}{\Omega_-} \right) , 
\end{equation}
where we have assumed $\Omega_+ > \Omega_- > 0$. Thus for our spectrum, for which the range is $[-\omega,\Omega]$, the Lamb shift is approximately  
\begin{equation}
     \Delta \omega_{\ms{L}} \approx (\gamma/2\pi) \ln(\Omega/\omega) . 
\end{equation}

\subsubsection{The effect of the finite separation between transitions: effective level coupling} 

As mentioned in Section~\ref{wellsep}, if two transitions are not well separated, so that the frequency difference between them, $\Delta\omega$, is not much larger than the decay rates of both transitions, then the transitions are effectively coupled together by their mutual coupling to the quasi-continuum. When the transitions are exactly degenerate the MEW still describes the evolution correctly, but for the intermediate regime in which $0 \ll \Delta\omega \sim \gamma$ no simple master equation has yet been obtained. (Degenerate transitions are coupled to the quasi-continuum in an identical way and as a result behave as a single transition: the resulting evolution is described in the MEW using a single transition operator that is the sum of their respective transition operators.) 

As an example we simulate in Fig.~\ref{fig5} the evolution of two transitions, and examine how this evolution changes as the difference between their respective frequencies is reduced. In the top (bottom) row of Fig.~\ref{fig5} we place all the initial population in the upper level of the transition with the higher (lower) frequency. When the levels are well separated each decays exactly as predicted by the MEW, but as the separation is reduced population flows between the two levels as they decay. When it is the transition with the higher frequency that is populated the coupling causes its decay to be inhibited, while the reverse is initially true when the other transition is populated. When the transitions are sufficiently close to being degenerate the difference in their behavior reduces as it must. As far as we are aware this regime has not been extensively explored, although some previous simulations have been presented in~\cite{Majenz13}.  

\section{The Random Matrix Model: state-independent rates and the cascade structure}
\label{RMM} 

To simulate the action of a thermal bath on a quantum system, a model must --- at the very least --- take every initial state and transform it into the Boltzmann state at temperature $T$ as $t\rightarrow\infty$. (Two caveats that we will always assume in the following are i) that the coupling to the thermal bath is sufficiently weak that it does not significantly change the energy levels of the system (note that the Boltzmann state is defined in terms of these levels, and thus assumes a sufficiently weak interaction), and ii) that thermalization will only occur if the operator by which the system couples to the bath couples enough of the systems eigenstates together so as to provide at least one path from any given state to every other.) 

Coupling a system to a bath that consists of a near-continuum of independent harmonic oscillators achieves the above requirement, so long as all the oscillators are initially prepared in thermal states themselves. Moreover, such a bath is more than just a model, since it corresponds to the real bath of electromagnetic modes to which an atom couples in free space. This immediately implies to things: i) to the extent that real many-body baths possess universal behavior, the oscillator bath must correctly describe this behavior; ii) if a model does not reproduce the behavior of an oscillator bath, then there is at least one real bath that the model fails to describe. For this reason one expects bath models to reproduce the basic behavior of the oscillator bath. In particular that it produce a simple set of rate equations that i) are independent of the initial state of the system, and ii) relax the system to the Boltzmann state for any initial state. Note that the oscillator bath is not the only real many-body bath for which a master equation has been derived. The collision model is another, and this also shares the above properties of the oscillator bath, providing the support for the hypothesis that this behavior is universal~\cite{Hornberger06, Hornberger07b}. We also note that the desire for concrete models is motivated by exploring the behavior of open systems outside the regime in which simple master equations suffice.


The ``random matrix model'' (RMM) of a quantum bath is constructed as follows~\cite{Massimiliano03, Massimiliano03b, Lebowitz04, Gemmer06, Breuer06, Bartsch08, Silvestri14}. One defines as the ``bath'' a system with $N$ closely-spaced energy levels, and chooses these $N$ levels so that, when arranged in order of increasing energy, their density with respect to energy increases exponentially with energy. The interaction between the system and bath is chosen to have the form $XY$, in which $X$ is any operator of the system, and $Y$ is an operator of the bath that is a random matrix when written in the basis of its energy eigenstates states. The random matrix is defined in the standard manner of random matrix theory~\cite{Guhr98, Mehta91, Mehta67, Porter65}. This model of a thermal bath has a number of compelling features. First, since each system transition is explicitly coupled to a continuum it is immediate that the model generates rate equations as per Fermi's golden rule. Since the resulting rates are proportional to the density of states at the destination, the exponential increase in this density provides the correct ratio's of upward and downward rates to generate the Boltzmann steady-state, and the temperature comes directly from the exponent chosen for the  exponential density. The fact that real many body systems have a density of states that increases exponentially, and that it is precisely this exponent that sets the temperature is one of the models compelling aspects. Another is that the random interaction matrix reflects the fact that the eigenstates of real, thermalizing many-body systems have the typicality property of randomly sampled states~\cite{Popescu06, Goldstein06, Rigol08}. Finally, the bath thermalizes the system without requiring that it is itself placed in a thermal state. The initial state of the bath may be chosen to be pure, and distributed uniformly over a sufficiently narrow energy window. This also reflects the state of real thermalizing many-body baths, and as such generates thermalization from first principles. 

The RMM is interesting because it explicitly incorporates many features of real many-body systems, and does so in a simple way. Nevertheless, as we now show, it fails to reproduce one important property of the MEW, the fact that the MEW's transition rates are independent of the initial state of the system. The reason for this failure is quite simple, and is a consequence of two things: i) to obtain the correct ratio of upward and downward transition rates, these rates are made to depend on the bath energy at the transition destination (by making the density of states depend on energy), and ii) the upward and downward transitions are mediated by coupling to the same bath states. Because the total energy of the system and bath are approximately conserved by the evolution, a necessary result of conditions i) and ii) is that the rate for a given transition (upward or downward) depends on the initial energy of the system (and thus on the initial state of the system), since this determines the energy the bath will have at the destination of any transition. This fact is most easily seen from a simple example. 


Let us denote the initial energy of the bath by $E_{\ms{bath}}$, and consider a single transition of the system in which the upper level, $|\mbox{e}\rangle$, has energy $\Delta E$ and the lower level, $|\mbox{g}\rangle$, has energy 0. If the system starts in the upper level, then when making the transition to the lower level the energy of the bath at the destination is $E_{\ms{dest}} = E_{\ms{bath}} + \Delta E$. Now consider what happens when the system starts in the lower level and reaches $|\mbox{e}\rangle$ by making the transition in the upward direction. In this case the total energy of the system and bath is $E_{\ms{Tot}} = E_{\ms{bath}}$. The system is now in state $|\mbox{e}\rangle$ and is ready to make the downward transition. But the energy of the bath at the destination for this transition is merely $E_{\ms{bath}}$. Thus when the system starts in $|\mbox{e}\rangle$ the transition rate for downward transitions is higher than when the system starts in $|\mbox{g}\rangle$.   

From our analysis in Section~\ref{struct} we see that the oscillator bath avoids the above problem because the quasi-continua that couple upward transitions are not the same as those that couple downward transitions. This allows the bath to set the ratio of upward and downward rates independently of the energy of the bath. To modify the random matrix model so that different quasi-continua mediate the upward and downward transitions would require a structure with a complexity similar to that of the oscillator bath. To see this note that whenever the system undergoes a downward transition the final states of that transition become the initial states for the upward transition. One cannot couple these initial states for the upward transition back to those for the downward transition because then the upward and downward transitions are coupled to the same continuum. One must use new quasi-continua for the destination of the each traversal of the transition, which results in the cascade structure. 

Since the oscillator bath results in a master equation that is independent of the initial state of the system, it is reasonable to assume that generic many-body thermal baths also have this property. This implies that the RMM is missing one of the key properties of many body systems, and that to extend the RMM to include this property  requires adding at least as much complexity as that of the cascade structure of the oscillator bath. If one wishes to use the RMM as method for numerical simulation of open systems, this issue can be partially circumvented by using different initial states of the bath for different initial state of the system, so as to maintain the same initial total energy. However, this is only effective in restricted settings: it is not possible for simulations of systems with time-dependent Hamiltonians since the total energy changes with the Hamiltonian. We can conclude that RMM's are only capable of simulating a restricted class of open systems. 

\vspace{3mm}
\section{Conclusion}
\label{conc}
\vspace{-2mm}

We have presented an analysis of the oscillator bath in which it is decomposed into a network couplings that consists of a cascade of the quasi-continuum coupling structures of Fermi's golden rule. This picture of the oscillator bath facilitates a comparison between the scenarios of Ferm's golden rule, coupling to an oscillator bath, and the random matrix model of a thermal bath. This analysis makes it intuitively clear how the rates of the Markovian master equation emerge as a result of a cascade of couplings to quasi-continua, as well as revealing precisely the relationship between the oscillator bath model and the random matrix model. We have shown that the random matrix model fails to reproduce one important property of the oscillator bath (assumed to be a  property of generic many-body baths), namely that the induced master equations are independent of the initial energy of the system. The analysis also makes it clear that in order to generate state-independent master equations an extension of the random matrix model would require at least as much complexity as the oscillator bath. 

We have used the full solution to the scenario of Fermi's golden rule to examine how the evolution induced by an oscillator bath deviates from the master equation due to the finite values of the bath cut-off frequency $\Omega$ and the transition frequency $\omega$. We have also examined, via simulations, the deviation due to the finite value of the difference between the frequencies of non-degenerate transitions, $\Delta \omega$. As a topic for future research, it may be interesting to ask whether a generalized version of Fermi's golden rule with two initial states might provide a route to obtaining simple equations describing transitions separated by frequencies on the order of the decay rates, a regime outside of current master equation techniques. 

Finally, we noted that the oscillator bath is able to generate thermal steady-states only because the coupling between the system and each of the bath oscillators is chosen to be proportional to the oscillator quadratures. There is no such requirement in coupling a system to a more generic many-body bath, since it can be expected to thermalize the system so long as it thermalizes its many constituents. As a topic for future research, it may be interesting to ask whether an analysis in terms of cascade structures might provide insight into the behavior of more generic bath models involving nonlinear systems, and the conditions under which such models induce thermal behavior. 

\textit{Acknowledgements:} KJ would like to thank Jordan Horowitz for helpful discussions. 


%

\end{document}